\documentclass[final]{siamltex}

\usepackage{amssymb}
\usepackage[english]{babel}
\usepackage{graphicx}
\usepackage{textcomp}
\usepackage{amsmath}

\numberwithin{equation}{section}

\title{On trajectories of  vortices in the compressible fluid on a two-dimensional manifold
\thanks{This work was supported by National
Science Council in Taiwan under Grants Nos. NSC
96-2911-M-001-003-MY3 and NSC 99-2115-M-126-001, by National Center
for Theoretical Sciences in Taiwan and Academia Sinica (Taiwan)}}

\author{Olga S. Rozanova\thanks{Department of Mechanics and Mathematrics, Moscow
State University, Moscow 119992 Russia ({\tt
rozanova@mech.math.msu.su})}
        \and Jui-Ling Yu\thanks{Department of Financial and Computational Mathematics, Providence
University, Taichung, 43301, Taiwan; and National Center of
Theoretical Sciences at Taipei, Physics Division, National Taiwan
University, Taipei 10617, Taiwan}\and Chin-Kun Hu\thanks{Institute
of Physics, Academia Sinica, Nankang, Taipei 11529, Taiwan and
Center for Nonlinear and Complex Systems and Department of Physics,
Chung Yuan Christian University, Chungli 32023, Taiwan}}

\begin{document}

\maketitle

\begin{abstract}
For the model of a compressible barotropic fluid  on a two
dimensional  rotating Riemmanian manifold we discuss a special class
of smooth solutions having a form of  a steady non-singular vortex
moving with a bearing field. The model can be obtained from the
system of primitive equations governing the motion of air over the
Earth surface after averaging over the height and therefore the
solution obtained can be interpreted as a tropical cyclone which is
known as a long time existing stable vortex. We consider
approximations of $l$- plane and $\beta$ - plane used in geophysics
for modeling of middle scale processes  and equations on the whole
sphere as well. We show that the solutions of the mentioned form
satisfy the equations of the model either exactly or with a
discrepancy which is small in a neighborhood of the trajectory of
the center of vortex. We perform a numeric study of the change of
the shape of the vortex affected by the neglecting the discrepancy
term.
\end{abstract}




\pagestyle{myheadings} \thispagestyle{plain} \markboth{O. ROZANOVA,
J.-L.YU AND  C.-K. HU}{TRAJECTORIES OF VORTICES IN THE COMPRESSIBLE
FLUID ON A 2D MANIFOLD}



\section{Introduction}

A lot of physical models can be reduced to the equations of motion
of compressible medium  on a two-dimensional manifold. One of the
most important is the model of  dynamics of the atmosphere. The
vertical scale of the atmosphere is much more larger than the
horizontal one, therefore one can reduce the initial primitive
three-dimensional system of equations to a two-dimensional one,
convenient for describing many kinds of motions of middle and global
scale, using a special procedure of averaging over the height.

There are a lot of papers  where the equations governing the
atmospheric motion are reduced to the two-dimensional system of
incompressible Navier-Stokes equations with viscosity (see
\cite{Temam},\cite{DKM}, and references therein), on this way a lot
of results on existence and uniqueness of solution to the Cauchy
problem  were obtained. The effect of compressibility makes the
situation more difficult and the results on the global in time
correctness for the viscous case are very scared (a recent state of
art can be found in \cite{Feireisl}).
 If we neglect the viscosity as they usually
do in the meteorology  (thus we get the quasilinear hyperbolic
system for compressible fluid), we find ourself in a more
complicated situation since the solution to the Cauchy problem
basically loses its initial smoothness with time. Recent
developments in the study of the Cauchy problem for the  equations
for compressible fluids are reviewed in \cite{ChenWang}.

Nevertheless, it is well known that  there exist very stable
vortices in the atmosphere that move without changing their  shape
for  days and even weeks. Most notorious example is the tropical
typhoon, one of the most devastating weather phenomena in the world.
In practice, the most important problem is to describe (or better
predict) the path of typhoon, whereas the processes inside the
vortex are of theoretical interest.

In our earlier papers \cite{OR2004}, \cite{OR_JLY_CKH_2010} we
developed a theory on possible trajectories of a stable vortex with
a linear profile of velocity governed by the system of equations of
compressible fluid on a plane (see Sec.\ref{4.1} for details). These
vortices can be interpreted as tropical cyclones in the conservative
phase of its dynamics, thought they have a number of nonrealistic
features. First, the components of the velocity  and pressure rise
unboundedly as the distance from the center increases (we call these
vortices non-localized). Second, the curvature of the Earth surface
was not taken into account even in a simplest form. Nevertheless, as
it was shown in \cite{OR_JLY_CKH_2010} by comparing with
observational data, the theoretical trajectories of vortices can
imitate the real tropical cyclone paths.  In fact, they are a
superposition of two circular motions and correspond to  the natural
circular (or rather parabolic, taking into account the change of the
Coriolis parameter with the latitude) trajectories, loops, reversal
points, etc. Moreover, the explicit expression for the trajectory
was  obtained.

In the present paper we show that the trajectories of localized
vortices that model the physical process more adequately are very
similar to the trajectories of the non-localized vortex with linear
profile of velocity  considered in \cite{OR_JLY_CKH_2010} under some
realistic assumptions on the background field of pressure.

It is worth mentioning that the exact form of equations describing
the atmosphere is not known. More exactly, in the primitive system
of equations  we can take into account some addition forces or
sources of energy, e.g. due to phase transitions, and the exact form
of these terms  can hardly be found,  particulary after the
averaging procedure.

We will base on the assumption on existence of special form of
solution for the system of equations of the  atmosphere dynamics
(steady vortex moving in a bearing field) together with the
assumption on a smallness of  the source term that guarantees the
existence of this special solution, at least in a domain of space
that we are interesting to control.

We can also look at the problem from another side and  call a
vector-function the "$\delta$ - approximate" solution to a certain
system in a domain $D\in {\mathbb R}^2\times {\mathbb R}_+$ if the
discrepancy term $Q$ arising after substitution of this function to
the system is less than $\delta$ in the uniform norm. Thus, we can
consider the term $Q$ either as some source term that guarantees the
existence of the {\it exact}  vortex solution or as a {\it
discrepancy}, in the latter case the vortex solution can be
considered as an approximative  one.

The paper is organized as follows. In Section \ref{Sec2}, we recall
the derivation of a bidimensional model of the atmosphere dynamics
based on the primitive  equations for compressible viscous
heat-conductive gas in the physical 3D space. In Section \ref{Sec3},
we develop a general approach to the solution in the form of
"frozen" vortex moving in the exterior field. In Section \ref{Sec4}
we demonstrate this method for the vortex solutions in the $l$ --
plane and the $\beta$ -- plane models and consider different shape
of possible vortices (both localized and non-localized). Moreover,
we compare the trajectories obtained here with the the trajectory of
vortex with linear velocity profile where the exact analytical
result is available. Further, we perform direct numerical
computation of the moving vortex in the 2D models of the $l$-plane
and $\beta$-plane and compare the position of the center of vortex
with the result obtained by our method. In Section \ref{Sec5} we
consider the case of a sphere.  We construct global exact solutions
for non-rotating sphere  and study the trajectories of approximative
vortices for the rotating case as well.
 Finally, a discussion about our
topics  is provided in Section \ref{Sec6}.

\section{Bidimensional models of the atmosphere
dynamics}\label{Sec2}

We do not dwell at the procedure of averaging and refer to our paper
\cite{OR_JLY_CKH_2010}, where we use the approach by \cite{Obukhov}
and \cite{Alishaev}. The initial (primitive) three-dimensional
system relates to the motion of compressible rotating, viscous,
heat-conductive, Newtonian polytropic gas \cite{Landau}, the
resulting two-dimensional system consists of three equations for
density $\varrho(t,{ x}),$ velocity ${\bf U}(t,{ x})$ and  pressure
$P(t,{ x})$  (the equation for the velocity is vectorial):
\begin{equation}\label{2d_U}
 \varrho(\partial_t {\bf U} +  ({\bf U} \cdot \nabla ){\bf U} + l L
{\bf U}) + \nabla P = F_1,
\end{equation}
\begin{equation}\label{2d_rho}
\partial_t \varrho +  \nabla \cdot ( \varrho {\bf U}) =0,
\end{equation}
\begin{equation}\label{2d_p}
\partial_t P + ({\bf U},\nabla P)+\gamma P\,{\rm div}{\bf
U}=F_2,
\end{equation}
where $F_1$ and $F_2$ are some source terms (see
\cite{OR_JLY_CKH_2010}) for details, $\quad L = \left(\begin{array}{cr} 0 & -1 \\
1 & 0
\end{array}\right)$, $l$ if the Coriolis parameter, a smooth function of $x$, its concrete form we
discuss below. In fact, system \eqref{2d_U} -- \eqref{2d_p} consists
of equations of balance  of mass, momentum and total energy, $t\in
\mathbb R_+$, $x\in \mathcal M$, where $\mathcal M$ is a 2D
Riemmanian manifold.

We can consider the system   at any 2D Riemannian manifold $\mathcal
M$ (with or without boundary).  Thus, we have to take into account
the curvilinear metrics of the space and write the system in the
respective curvilinear coordinates. In this case we mean by partial
derivatives the covariant derivatives \cite{Diff_Geom}. In
particular, to model the motion of planetary scale we should use the
 spherical coordinates (\cite{Pedloski}). Nevertheless, for the
phenomena of  middle scale it is convenient to use the approximation
of plane, without taking into account the effects of curvature.

Now we recall shortly the procedure of averaging. Let $\rho, \,
u=(u_1,u_2,u_3),\, p, \,$ denote in the {\it three-dimensional}
density, velocity and pressure. Namely, all these functions depend
on $(t,x,z),$ $z\in\mathbb R_+$. Let us introduce $\hat\phi$ and
$\bar f$ to represent for taking the average of $\phi$ and $f$ over
the height, respectively. The averaged values are introduced as
follows: $\displaystyle\hat\phi :=\int_0^\infty \phi\,dz,\quad \bar
f :=\frac{1}{\hat\rho}\int_0^\infty \rho f\,dz$, where $\phi$ and
$f$ are arbitrary functions, and denote $\varrho(t,{ x})=\hat\rho,
\,P(t,{ x})=\hat p,\,{\bf U}(t,{x})=(\bar u_1, \bar u_2).$ Moreover,
the usual adiabatic exponent, $\tilde\gamma$, is related to the
``two-dimensional" adiabatic exponent $\gamma$ as follows:
$\gamma=\displaystyle\frac{2\tilde\gamma-1}{\tilde\gamma}<\tilde\gamma.$

 The impenetrability conditions are included in the model. These conditions
ensure that the derivatives of the velocity equal to zero on the
Earth surface and a sufficiently rapid decay for all thermodynamic
quantities as the vertical coordinate $z$ approaches to infinity. In
other words, the impenetrability conditions make sure the
boundedness of the mass, energy, and momentum in the air column.
They also provide the necessary conditions for the convergence of
integrals.

The source terms $F_1$ and $F_2$ include basically  the (turbulent)
viscosity, heat conductivity, and may be something else. In fact,
the form of $F_1$ and $F_2$ for real geophysical model depends on
the model considered, in any case these terms  are believed to be
small and to have a regularizing effect. Indeed,  the solution to
system \eqref{2d_U}--\eqref{2d_p} (i.e. the compressible Euler
system) generally loses smoothness within a finite time with
formation of shocks, and it needs to introduce some smoothing terms
to avoid this phenomenon not inherent to the atmospheric motion.

For  three-dimensional equations it is the practice to introduce the
entropy function $s$, which is connected with the pressure and
density through  the state equation
\begin{equation*}\label{st_eq_3D}
 p=a\,e^s \rho^{\tilde\gamma},
\end{equation*}
with  a constant $a>0$.

 We can also introduce  the 2D entropy $S(t,{x})$ connecting with $\varrho$ and
$P$ similarly to \eqref{st_eq_3D}, where we use $\gamma $ instead of
$\tilde\gamma.$ Then instead of equation \eqref{2d_p} we get the
equation for the entropy
\begin{equation}\label{2d_s}\partial_t S +({\bf U},\nabla S)=F_3,
\end{equation}
with a source $F_3$, see \cite{OR_JLY_CKH_2010} for details.

For the sake of simplicity and to make our idea clear in this paper
we will consider only the barotropic case of constant entropy $S_0$,
in other words, the state equation is \begin{equation} P=C
\rho^\gamma,\gamma>1, \qquad C={\rm const} >0,\end{equation} and
\eqref{2d_s} is assumed to hold identically with $F_3=0$.

Thus, the system under consideration is reduced to two equations
\eqref{2d_U}, \eqref{2d_rho}.

It will be convenient to  introduce a new variable
$\pi=P^{\frac{\gamma-1}{\gamma}}.$
 For the new unknown variables $\pi(t,x), {\bf U}(t,x)$
we obtain the system
\begin{equation}\label{2Dsys}\begin{aligned}
\partial_t {\bf U} +  ({\bf U} \cdot \nabla ){\bf U} +
l L {\bf U} + c_0\,\nabla
\pi =F,\\
\partial_t \pi +  (\nabla \pi \cdot  {\bf U}) +
(\gamma-1) \pi \, {\rm div}  {\bf U}=0, \end{aligned}
\end{equation}
with a new source term $F(t,x)$ and $c_0= \frac{\gamma}{\gamma-1}
C^{\frac{1}{\gamma}}.$

\section{"Frozen" vortex: a general approach}\label{Sec3}

Let us change the coordinate system in such a way that the origin of
the new system ${\bf x}=(x_1,x_2)$ is located at a point ${\bf
X}(t)=(X_1(t),X_2(t))$ (here and below we use the bold font for $\bf
x$ to denote the local coordinate system). Now ${\bf U}= {\bf
u}+{\bf V},$ where
 ${\bf V}(t)=(V_1(t),V_2(t))=(\dot X_1(t),\dot X_2(t))$.
Thus, we obtain a new system
\begin{equation}\begin{aligned} \label{2d_sh_u}
\partial_t {\bf u} + ({\bf u} \cdot \nabla ){\bf u} +
\dot {\bf V}+ l({\bf X}(t)+{\bf x}) \, L \,({\bf u}+{\bf V}) +c_0\,
\nabla \pi = F,\end{aligned}
\end{equation}
\begin{equation}\label{2d_sh_pi}
\partial_t \pi +  \nabla \pi\cdot {\bf u} + (\gamma-1)\,\pi\,{\rm div}\,{\bf
u}\, =\,0.  \end{equation}
Given a vector ${\bf V}$, the trajectory
can be found by integrating the system
\begin{equation}\label{3.4}\dot X_1(t)=V_1(t),\qquad \dot
X_2(t)=V_2(t).\end{equation}

Below we  look for a solution with special properties, namely,
 a steady divergency free
vortex "frozen" into a certain exterior pressure field. Together
with this we study {\it a class of sources $F$ allowing the
existence of such solutions.} Let us stress that we do not prescribe
specific boundary conditions and  only {\it assume} that
$${\bf u} \mbox{\quad does not depend of}\quad
t,\quad {\bf u}(0)=0 \mbox{\quad and \quad} {\rm div \,\bf u} =0.$$
This means that there exists a potential $\Phi(x_1,x_2)$ such that
\begin{equation}\label{u}{\bf u}=\nabla_\bot \Phi = (\Phi_{x_2}, - \Phi_{x_1})).\end{equation} Thus,
the velocity field satisfies the Cauchy-Riemann conditions
$$\partial_{x_1}{u_1}= \partial_{x_2}{u_2}, \qquad \partial_{x_2}{u_1}=
-\partial_{x_1}{u_2}.$$ In \cite{DST} it is proved that for the case
of $l$-plane  this condition is necessary for the existence of a
stable vortex.

Further, we {\it assume} that
\begin{equation}\label{pi_0}\pi=\pi_0(x_1,x_2)+\pi_1(t,x_1,x_2),\end{equation}
 where
\begin{equation}\label{constraint}(\nabla\pi_0, \nabla_\bot \Phi) = 0.\end{equation}
We call the time-dependent component $\pi_1$  {\it a bearing field
of pressure}. It will be  important for us that the gradient of the
bearing field is rather small. The time-independent part of pressure
$\pi_0$ relates to the vortex itself.

Let $A$ be an appropriate ($2\times 2$) matrix which depends on the
situation to be discussed below. We consider the following equation
for the potential of velocity $\Phi$ and the steady part of pressure
$\pi_0$:
\begin{equation}\begin{aligned} \label{Phi_pi_0}
(\nabla_\bot \Phi, \nabla)\nabla_\bot \Phi + A \nabla_\bot \Phi +
c_0 \nabla \pi_0=0.\end{aligned}
\end{equation}

Let us suppose that we succeed to solve system  \eqref{Phi_pi_0},
\eqref{constraint}. Then from \eqref{2d_sh_pi} we get a  linear
equation for $\pi_1$:
\begin{equation}\label{pi1}\begin{aligned}
\partial_t \pi_1 +  \nabla \pi_1\cdot {\bf u}= 0,
\end{aligned}
\end{equation}
which  can be solved for any initial condition $\pi_1(0,{\bf x})$.

Further, from  \eqref{2d_sh_u} we obtain
\begin{equation}\begin{aligned} \label{v}
\dot {\bf V}(t)+ l({\bf X}(t)+{\bf x}) L {\bf V}(t) +(l({\bf
X}(t)+{\bf x})L-A){\bf u}({\bf x}) +c_0 \nabla \pi_1(t,{\bf x})=
F,\end{aligned}
\end{equation}
or
\begin{equation}\begin{aligned} \label{x}
\ddot {\bf X}(t)+ l({\bf X}(t)+{\bf x}) L \dot{\bf X}(t) +(l({\bf
X}(t)+{\bf x})L-A){\bf u}({\bf x})+c_0 \nabla \pi_1(t,{\bf x})=
F.\end{aligned}
\end{equation}
Let us denote
\begin{equation}\begin{aligned}\label{Q}
Q =-(l({\bf X}(t))-l({\bf X}(t)+{\bf x})) L \dot{\bf X}(t)
+\\(l({\bf X}(t)+{\bf x})\,L-A){\bf u}({\bf x})-c_0\left[\nabla
\pi_1(t,{\bf x})\Big|_{{\bf x}=0}- \nabla\pi_1(t,{\bf
x})\right].\end{aligned}\end{equation}

From \eqref{x} and \eqref{Q}  we get
\begin{equation}\begin{aligned} \label{xQ}
\ddot {\bf X}(t)+ l({\bf X}(t)) L \dot{\bf X}(t)+c_0 \nabla
\pi_1(t,{\bf x})\Big|_{{\bf x}=0}= F - Q.\end{aligned}
\end{equation}
If  $F$ equals  $Q$, then the position of the center of vortex under
consideration can be found from the following equation:
\begin{equation}\begin{aligned} \label{xQ_exact}
\ddot {\bf X}(t)+ l({\bf X}(t)) L \dot{\bf X}(t) +c_0 \nabla
\pi_1(t,{\bf x})\Big|_{{\bf x}=0}= 0.\end{aligned}
\end{equation}

Thus,  \eqref{2d_sh_u} can be rewritten as
\begin{equation}
\begin{aligned} \label{separation}
({\bf u} \cdot \nabla ){\bf u} + A{\bf u}+c_0 \nabla \pi_0+\\
\ddot {\bf X}(t)+ l({\bf X}(t)) L \dot{\bf X}(t) +c \nabla \pi_1(t,{\bf x})\Big|_{{\bf x}=0}= \\
F-Q,\end{aligned}
\end{equation}
the terms in the first row depend only on the space variable ${\bf
x}$, the terms in the second row depend only on $t$, therefore for
the case $F-Q=0$ we obtain a complete separation of variables for
solution satisfying \eqref{2Dsys}, \eqref{u}, \eqref{pi_0},
\eqref{constraint},  \eqref{pi1}, \eqref{xQ_exact}.

If $F-Q$ is not zero, however it is  small (for example, the uniform
norm $\|F-Q\|<\delta$), we can talk about a "$\delta$- approximate"
separation of variables, $F-Q$ plays a role of the discrepancy.

\begin{definition}
We call a couple $(\pi(t,x), {\bf U}(t,x))$ the $\delta$ -
approximate solution to system \eqref{2Dsys} in a domain
$D(\delta)\in \mathbb R_+\times {\mathcal M}$, if $(\pi, {\bf U})$
satisfies \eqref{2Dsys} with a discrepancy, whose uniform norm is
smaller than $\delta $ for any $(t,x)\in D(\delta)$.
\end{definition}

In fact we have to choose the matrix $A$ in such a way as to ensure
the smallness of the discrepancy term $F-Q$. Since we are interested
in studying the behavior of solution near the origin of the moving
coordinates system,  we expand in \eqref{Q} the {\it known}
functions $l({\bf x})$, ${\bf u}({\bf x})$ and $\pi_1(t,{\bf x})$ in
the origin and get $Q=O(|{\bf x}|)$ provided $\dot {\bf X}(t)$ keeps
boundedness. Basically a solution to the nonlinear equation
\eqref{xQ} blows up in a finite time $t_*$, this entails an
unboundness of $\dot {\bf X}(t)$. Thus, our considerations are valid
for $t<t_*$, until the formation of the shock wave in a neighborhood
of the  trajectory ${\bf X}(t)$ of the moving coordinate system.

Let us list once more the steps of our method. First we should solve
 \eqref{constraint}, \eqref{Phi_pi_0}, then set an initial distribution
for the bearing pressure field  $\pi_{1}(0,x_1,x_2),$ then find the
respective time-dependent  pressure component $\pi_1(t,x_1,x_2)$ and
at last integrate \eqref{x} to find exact (for $F=Q$) or approximate
(for small $F-Q$) trajectory of the exact (or $\delta$- approximate)
vortex solution.

Thus, the main problem is to solve the system \eqref{constraint},
\eqref{Phi_pi_0},
 with respect to unknown scalar functions $\Phi$ and
$\pi_0$ (may be under suitable boundary conditions).

Let us find a necessary condition for the function $\Phi$ that
allows it to be a part of solution of \eqref{constraint},
\eqref{Phi_pi_0}. We take the inner product of \eqref{Phi_pi_0} and
$\nabla_\bot \Phi$ and get {\it the master equation}
\begin{equation}\label{master}
 \nabla_{x_1 x_2} \Phi \,( (\nabla_{x_1} \Phi)^2-(\nabla_{x_2}
 \Phi)^2)\,+\,(A\nabla_\bot\Phi,\nabla_\bot\Phi)
  =\,
\nabla_{x_1} \Phi\,\nabla_{x_2} \Phi\,( \nabla_{x_2 x_2} \Phi-
\nabla_{x_1 x_1} \Phi).
\end{equation}
If the solution of \eqref{master} satisfies the identity
\begin{equation}\begin{aligned} \label{cond_pi}
\nabla \times ((\nabla_\bot \Phi, \nabla)\nabla_\bot \Phi + A
\nabla_\bot \Phi)=0,\end{aligned}
\end{equation}
then one can find $\pi_0$ such that the couple ($\Phi, \pi_0$)
solves \eqref{constraint}, \eqref{Phi_pi_0}.

Let us summarize the results of this section.

\begin{theorem}\label{T1}
Let $\Phi({\bf x})$ be a solution to equation \eqref{master},
satisfying condition \eqref{constraint} with a matrix $A$ with
smooth coefficients dependent only on space variables. Further, let
$\pi_0({\bf x})$ be a solution to \eqref{Phi_pi_0}, $\pi_1(t,{\bf
x})$ be a solution to \eqref{pi1} with $ {\bf u }= \nabla_\bot
\Phi$, $\, {\bf u }(0)= 0$, and ${\bf X}(t)$ be a solution to
\eqref{xQ_exact}. Then the couple $(\pi, {\bf U})$, where
\begin{equation}\label{sol}
\pi(t,x)=\pi_0(x-{\bf X}(t))+\pi_1(t,x-{\bf X}(t)),\quad {\bf
U}(t,x)=\dot{\bf X}(t)+{\bf u}(x-{\bf X}(t))
\end{equation}
solve system \eqref{2Dsys} with $F=Q$, $Q$ is given by \eqref{Q}.
\end{theorem}

\begin{theorem} \label{T2} We assume  $F=0$. Let $\Phi, \pi_0, \pi_1, {\bf X}(t)$ be as in Theorem \ref{T1},
moreover, they are classical solutions to the respective equations.
Assume $|\dot {\bf X}(t)|$ to be bounded for $t\le T$. Then there
exists a neighborhood $D(\delta)$ of the trajectory $x={\bf X}(t)$
such that \eqref{sol} is the $\delta$- approximate solution to
system \eqref{2Dsys} for $t\le T$. The dimensions of $D(\delta)$
depend on derivatives of known functions, namely, $|\nabla l(x)|$,
$|\nabla \pi(t,x)|$, $|{\bf h}|$, where ${\bf h}=(h_1,h_2)$,
$h_j=\sum\limits_{i=1}^2(lL-A)_{ij} u_i,\, j=1,2,$ ${\bf u}=(u_1,
u_2).$
\end{theorem}

{\em Proof}. The proof of Theorem \ref{T2} follows from  estimating
the term $Q$ given by \eqref{Q}.\endproof

Below we dwell on important for geophysics cases of application of
the method.

\section{Steady vortices on a plane}\label{Sec4}

The  "plane" models are used in meteorology to describe the
processes of small and medium scale, where the curvature of the
Earth surface does not play a crucial role and the change of the
Coriolis parameter can be modeled in a simplest way.

Let ${\bf x}_0=(x_{01},x_{02})$ be a point on the Earth surface,
$\varphi_0$ be the latitude of some fixed point ${\bf x}_0.$ The
simplest possible "plane" model is the so called $l$-plane model,
where the Coriolis parameter $l$ is treated as a constant: $\,l =
l_0=2 \Omega \sin \varphi_0$,  $\Omega $ is the vertical component
of the angular velocity of the Earth rotation. The model, that is
believed to be more adequate to describe  the weather processes, is
the $\beta$-plane model, where the Coriolis parameter $l$ is
approximated by a linear function:   $\,l=l_0+\beta x_2,$ the
constant $ \beta=\frac{2 \omega_3}{R} \cos \varphi_0,$ where $R$ is
the radius of the Earth. In fact, in the $l$-plane model we neglect
the parameter $\beta$, which is much more smaller than $l_0$. For
example, for $\varphi_0\,=\,30^\circ,\, l_0\,\approx\,7.3\times
10^{-5}\,{\rm s}^{-1}, \,\beta \,\approx\, 2\times 10^{-11} {\rm
s}^{-1} $, where we used the values  $R=6.4\times 10^6 \rm m$ and
$\omega_3=7.3\times 10^{-5} \rm rad/s$.

It can be readily checked that for $l=l_0 = \rm const$ every
potential
$$ \Phi=\Phi(r),\qquad r=\sqrt{x_1^2+x_2^2},
$$
solves the master equation \eqref{master} with $A=l_0 L$. We can get
different shapes  of vortices choosing different solutions.

For the $\beta$--plane model we will use the master equation with
the same matrix $A=l_0 L$. Recall that the choice of this matrix can
be different, nevertheless it is caused by a desire to make the
discrepancy term $F-Q$ as smaller as possible.

\subsection{Example 1: steady vortex on the $l$-- plane (linear profile of
velocity)}\label{4.1} We begin from the simplest case where for
$F=Q=0$ we have the complete separation of variables in equation
\eqref{separation}. Here we can obtain an exact  solution. Indeed,
if we choose $\Phi=\displaystyle \frac {b_0}{2} \, (x_1^2+x_2^2),\,
b_0=\rm const$, we get the velocity field with a linear profile
\begin{equation}\label{vel_lin_prof}
u_1\,=\,b_0 x_2,\quad u_2\,=\,-b_0 x_1.
\end{equation}
Further, if we choose the initial data for $\pi_1$ as a linear
function of the space variables,
$\,\pi_1(0,x)\,=\,M_{10}\,x_1\,+\,M_{20}\, x_2\, +\,K_0$,
$M_{10},\,M_{20},\,K_0= {\rm const}$, we get
\begin{equation}\label{pi_1_l_nl}
\pi_1(t,x)\,=\,M_1(t)\,x_1\,+\,M_2(t)\, x_2\, +\,K(t),
\end{equation}
where
\begin{equation}\label{M_0_N_0}
M_1(t)=M_{10} \cos b_0 t +M_{20} \sin b_0 t,\quad M_2(t)=M_{20} \cos
b_0 t -M_{10} \sin b_0 t.
\end{equation}
The trajectory obeys the equation
\begin{equation}\label{x_l_nonloc}\begin{aligned}
\ddot {\bf X}(t)+ l_0 L \dot{\bf X}(t) +c_0 {\bf
M}(t)=0,\end{aligned}
\end{equation}
where ${\bf M}(t)=(M_1(t), M_2(t))$.

The solution can be found explicitly:
\begin{align*}\label{exact_trajectory}
X_1(t)= & \quad X_1(0)+\frac{V_2(0)}{l}+\frac{c_0 M_{10}}{b_0 l}& &
\\+&\quad \left(\frac{V_1(0)}{l}-\frac{c_0 M_{20}}{l(b_0-l)}\right)\sin lt
-\left(\frac{V_2(0)}{l}+\frac{c_0 M_{10}}{l(b_0-l)}\right)\cos
lt&&\\+ & \quad \frac{c_0 M_{20}}{b_0(b_0-l)}\sin b_0t+\frac{c_0
M_{10}}{b_0(b_0-l)}\cos b_0t,&&\\X_2(t)= & \quad
X_2(0)-\frac{V_1(0)}{l}+\frac{c_0 M_{20}}{b_0 l}& &
\\+ &\quad\left(\frac{V_2(0)}{l}+\frac{c_0 M_{10}}{l(b_0-l)}\right)\sin lt
+\left(\frac{V_1(0)}{l}-\frac{c_0 M_{20}}{l(b_0-l)}\right)\cos
lt&&\\&\quad -\frac{c_0 M_{10}}{b_0(b_0-l)}\sin b_0t+\frac{c_0
M_{20}}{b_0(b_0-l)}\cos b_0t,&&\end{align*} (for $l\ne b_0$). This
is a superposition of two circular motions: the respective
frequencies are  $\frac{2\pi}{l_0}$ and $\frac{2\pi}{b_0}$.

On this way we obtain the same solution as in \cite{OR2004},
\cite{OR_JLY_CKH_2010}. The above formulae  were used in
\cite{OR_JLY_CKH_2010} to imitate and forecast the trajectories of
tropical cyclones based on real observational data. In fact, we are
going to show that a further sophistication of model do not implies
significant difference in the behavior of the vortex trajectory, at
least for the "realistic" (relative to meteorological data) values
of parameters.

\subsection{Example 2: approximately steady vortex on the $\beta$-- plane (linear profile of velocity)}

Let us choose again $A=l_0 L$, $\Phi=\displaystyle \frac {b_0}{2} \,
(x_1^2+x_2^2)\,$, $\,\pi_1(0,x)\,=\,M_{10}\,x_1\,+\,M_{20}\, x_2\,
+\,K_0$, $M_{10},\,M_{20},\,K_0= {\rm const}$. Thus,
\eqref{vel_lin_prof} gives the velocity field  and \eqref{pi_1_l_nl}
gives the background pressure field $\pi_1$  as in  Ex.1.

To obtain the solution of the form of the steady vortex we need to
set $F=Q$, where  $Q$ (see \eqref{Q}) has the form
\begin{equation}\begin{aligned}
Q =\beta\, x_2 \, L  \,\big(\dot{\bf X}(t)\, +\,{\bf u}({\bf
x})\big)\,+\,\beta \,X_2(t)\, L \,{\bf u}({\bf
x}).\end{aligned}\end{equation}

 To find the trajectory of the
steady vortex we have to integrate the equation \eqref{xQ}
\begin{equation}\label{tr_ex2}\begin{aligned}
\ddot {\bf X}(t)+ (l_0+\beta X_2(t))\, L \,\dot{\bf X}(t) +c_0\,
{\bf M}(t)=0.\end{aligned}
\end{equation}

If $F=0$, we can consider $Q$ as a vector of discrepancy, and due to
the smallness of $\beta$ for real meteorological models the value of
$Q$ is  essentially small until the blow up time of the respective
solution to the ODE \eqref{tr_ex2}.

\subsection{Example 3: steady localized vortex on the $l$-- plane}

\begin{figure}
\centerline{\includegraphics[width=0.6\columnwidth]{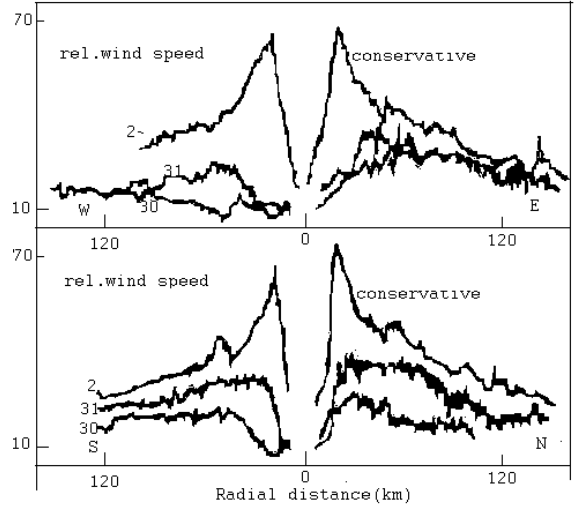}}
\caption{The tangential component of (relative) velocity in $m \cdot
s^{-1}$ on different stage of the typhoon formation (August 30,
August 31, September 2). The hurricane Anita, North Atlantic, 1977
\cite{IAV}. The horizontal scale is in kilometers.}\label{fig1}
\end{figure}

\begin{figure}
\centerline{\includegraphics[width=0.4\columnwidth]{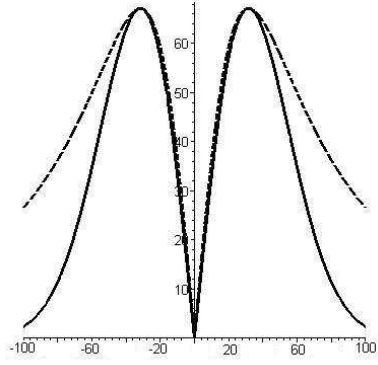}}
\caption{The tangential component of (relative) velocity in $m \cdot
s^{-1}$  for a localized vortex: exponential decay (solid line) and
power decay (dashed line). The horizontal scale is in kilometers.
}\label{PR_vel_gauss_power}
\end{figure}

As was mentioned before the velocity field \eqref{vel_lin_prof}
rises at infinity and is not realistic. Thus it is natural to
consider the vortex in some sense localized in the space.   We will
call the vortex "localized", if the velocity profile looks like the
observational data (e.g. \cite{IAV}, see Fig.\ref{fig1}). This kind
of initial data is used for the majority of numerical computations
simulating tropical cyclones. It is catching to select the form of
the vortex basing on the experimental data, for example, in the
class of potential functions
\begin{equation}\label{loc_power}
\Phi(r)=\frac{a}{(1+\sigma r^2)^{k}},\qquad a>0,\,\sigma>0,\,k>0.
\end{equation}
Indeed, the dashed line  in Fig.\ref{PR_vel_gauss_power} presents
the graph of the modulus of the radial component of the
corresponding velocity field for $k=\frac{1}{2},\quad a=7.8\times
10^{6}\,{\rm m^2/s} ,\quad \sigma= 5\times 10^{-10}\,{\rm m^{-2}}$.
The parameters give a correct maximum of velocity and a realistic
decay of the velocity at infinity. Nevertheless, one can remember
that in our model we use the averaged over the height velocity and
to construct the graph analogous to Fig.\ref{fig1} for averaged
velocity we need to analyze the experimental data for different
height levels (that we do not have at our disposal). Thus, the
profile of velocity for averaged data may be different from those
that presented in Fig.\ref{fig1}. At least, the value of amplitude
$a$ should be much more smaller due to the different direction of
rotation of the air for lower and upper parts of cyclone \cite{IAV}.
Moreover, if we use the pattern \eqref{loc_power}, we cannot  solve
explicitly equation \eqref{pi1} for the background pressure field.
Therefore we use a more localized (exponentially decaying) solution
to the master equation \eqref{master} with $A=l_0 L$:
\begin{equation}
\label{vel_gauss} \Phi=-B_0 \,{e^{-\frac{\sigma}{2}\, \left(
{x_1}^{2}+{x_2}^{2} \right) }},\end{equation} here $\,B_0$ is a
constant, $\,\sigma$ is a positive constant. The respective velocity
field
\begin{equation}\label{u_l_0}
u_1\,=\,B_0\,\sigma\,x_2\,e^{-\frac{\sigma}{2}\, \left(
{x_1}^{2}+{x_2}^{2} \right)},\qquad u_2\,=\,
-B_0\,\sigma\,x_1\,e^{-\frac{\sigma}{2}\, \left( {x_1}^{2}+{x_2}^{2}
\right)}
\end{equation}
vanishes at infinity and  has the same asymptotics as
\eqref{vel_lin_prof}  at the origin, with $b_0=B_0\,\sigma$. The
solid line in Fig.\ref{PR_vel_gauss_power} shows the modulus of the
radial component of  velocity for $B_0= 3.5\times 10^{-6}{\rm
m^2/s},\,\sigma=10^{-9}\,\rm m^{-2}. $

We are going to show that for the values of parameters  relevant in
the meteorology  such strict localization does nor lead to a big
difference in trajectories of the localized and non-localized
vortices.

It can be checked that condition \eqref{cond_pi} is satisfied and
therefore we can find the scalar function $\pi_0$. Namely,
\begin{equation}\label{pi_0_l_0}
\pi_0(x_1,x_2)\,=\,\frac{1}{c_0}\,\left(\frac{1}{2}\,B_0^2\,\sigma\,e^{-\sigma\,(x_1^2+x_2^2)}\,
-\,l_0\,B_0\,e^{-\frac{\sigma}{2}\,(x_1^2+x_2^2)}\right).
\end{equation}
Further, equation \eqref{pi1} can be explicitly solved, its general
solution is $$\pi_1={\it G} \left( {x_1}^{2}+{x_2}^{2},
t-\frac{1}{B_0\,\sigma}e^{\frac{\sigma}{2}\, ({x_1}^{2}+{x_2}^{2})}
\arctan \left(\frac {x_1}{x_2} \right) \right), $$ with any
differentiable function ${\it G}$. For example, for initial
background field
\begin{equation}\label{pi_1_l_0}
\pi_1(0,x_1,x_2)= R_0+\phi(x_1,x_2)\,(M_{10}\,x_1 +
M_{20}\,x_2\,+\,K_0),\quad
\end{equation}
$$R_0,\,M_{10},\,M_{10},\,K_0\,=\,const, \quad \phi_0(x_1,x_2)=e^{-{\sigma_0\,
({x_1}^{2}+{x_2}^{2})}},$$
we get
\begin{align*} \pi_1(t,x_1,x_2)=& &\\
R_0+\phi_0(x_1,x_2)\,\big(K_0\,+\,M_{10}\,(x_1\,\cos(
\phi(x_1,x_2)\,B_0\,\sigma\,t)\,
- \, & x_2\,\sin( \phi(x_1,x_2)\,B_0\,\sigma\,t)) +&\\
M_{20}\,(x_2\,\cos( \phi(x_1,x_2)\,B_0\,\sigma\,t)\, + \, &
x_1\,\sin( \phi(x_1,x_2)\,B_0\,\sigma\,t))\big),&
\end{align*}
with $\quad \phi(x_1,x_2)=e^{-\frac{\sigma}{2}\,
({x_1}^{2}+{x_2}^{2})}.$

The trajectory of the  vortex can be found by integration from
equation \eqref{xQ}:
\begin{equation}\label{x_l_loc}\begin{aligned}
\ddot {\bf X}(t)+ l_0 L \dot{\bf X}(t) +c_0\nabla \pi_1(t,{\bf
x})\Big|_{{\bf x}=0}=0.\end{aligned}
\end{equation}
It can be checked that $$\nabla \pi_1(t,{\bf x})\Big|_{{\bf
x}=0}=(M_1(t),M_2(t)),$$ with the same $M_1(t)$ and $M_2(t)$ as in
\eqref{M_0_N_0}.

Thus,  equation \eqref{x_l_nonloc} that governs the trajectory of
the non-localized vortex for this specific background pressure field
coincides with \eqref{x_l_loc}.

The discrepancy term (for $F=0$) here is
\begin{equation}\begin{aligned}
Q= c_0\left[\nabla \pi_1(t,{\bf x})-\nabla \pi_1(t,{\bf
x})\Big|_{{\bf x}=0}\right],\end{aligned}
\end{equation}
it depends only on the properties of the background pressure field
$\pi_1$. From the explicit expression for $\pi_1$, presented above,
one can see that $Q$ is bounded  and its norm vanishes at every
point of plane as the parameters of initial slope of the background
pressure field $M_{10}$ and $M_{20}$ tend to zero.

\subsection{Example 4: steady localized vortex on the $\beta$-- plane}

We use in this case the same master equation as in  Ex.3 and its
solution \eqref{vel_gauss}. Moreover, we use the same initial
background pressure field $\pi_1(0,{\bf x})$ as in Ex.3.
 The equation \eqref{xQ} for the  trajectory of the steady vortex is
\begin{equation}\label{beta_position}\begin{aligned}
\ddot {\bf X}(t)+ (l+\beta X_2(t)) L \dot{\bf X}(t) +c_0\nabla
\pi_1(t,{\bf x})\Big|_{{\bf x=0}}=0,\end{aligned}
\end{equation}
this is the same equation as \eqref{tr_ex2}, the discrepancy $Q$
(for $F=0$) is
\begin{equation*}\begin{aligned}
Q = \beta x_2 L  \,\big(\dot{\bf X}(t) +{\bf u}({\bf x})\big)+\beta
X_2(t) L {\bf u}({\bf x}) +
\\c_0\left[\nabla \pi_1(t,{\bf x})-\nabla \pi_1(t,{\bf
x})\Big|_{x=0}\right].\end{aligned}\end{equation*} It can be
considered as a small one due to the reasons described in Examples 2
and~3.

\subsection{Comparison of trajectories for the $l$  -- plane and
the $\beta$ -- plane models}\label{Comparison of trajectories}

We compared the trajectories of vortices corresponding to
approximate solutions
 under values of parameters close to the realistic ones. Namely,
we set $\Omega= 7.3\times 10^{-5}\rm s^{-1},$ the geographical
latitude $\phi_0=30^{\circ},$  the Earth radius $R=6.39\times 10^{6}
\rm m$, $c_0=0.1\, $(appropriate dimension), $B_0=3.5 \times 10^3
\rm m^2/s $, $\sigma=10^{-9}\rm m^{-2}$, $\sigma_0=8\times
10^{-13}\rm m^{-2},$ $M_0=2\times 10^{-3} \rm s^{-1},$
$N_0=10^{-3}\rm s^{-1},$ $X_1(0)=X_2(0)=0,\,$ $V_1(0)=-1\, \rm m/s,$
$V_2(0)=1\,\rm m/s$.

\begin{figure}
\centerline{\includegraphics[width=0.4\columnwidth]{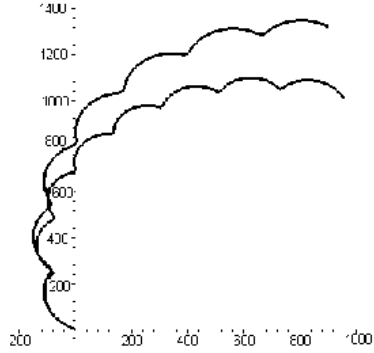}}
\caption{Trajectories of  vortices on $l$-plane (upper curve) and
$\beta$ - plane for 7 days period. The space scale is in
kilometers.} \label{l_b}
\end{figure}

 The result is the following: the difference between the $l$  -- plane and
the $\beta$ -- plane models
 is very small for these parameters up
to 2 days period of computation, further the difference increases,
however the behavior of the both trajectories remains similar (see
Fig.\ref{l_b}).

\subsection{Computational experiments with localized vortices on the plane: influence of the discrepancy term}

As it was shown before, if the discrepancy term is zero, the vortex
is steady and it moves according to \eqref{xQ}. Nevertheless, let us
study the influence of the discrepancy on the shape of vortex. It is
natural to expect that a big discrepancy destroys the vortex,
whereas a negligible one only modifies it.

\begin{figure}[h]
\begin{minipage}{0.5\columnwidth}
\centerline{\includegraphics[width=1.3\columnwidth]{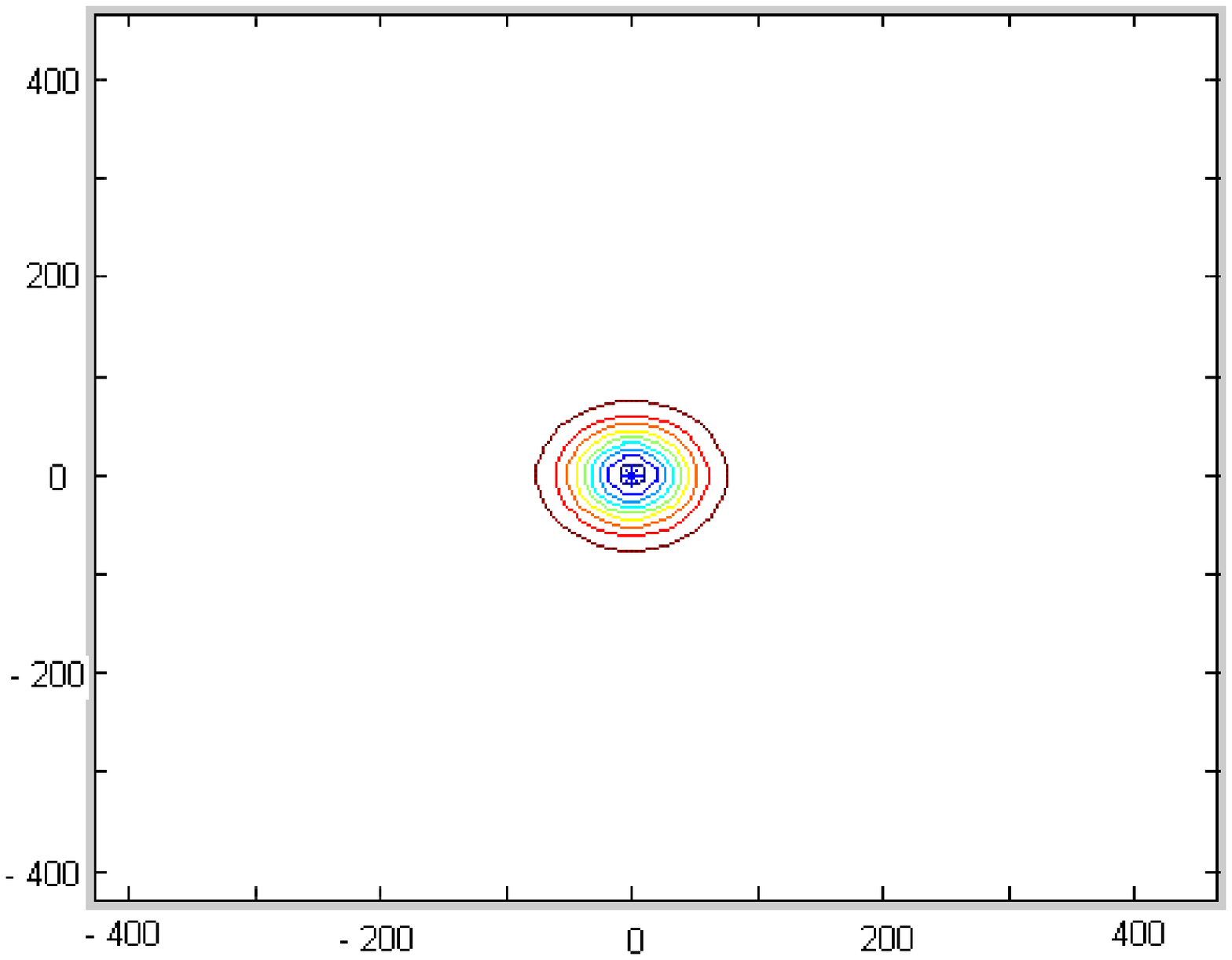}}
\end{minipage}
\begin{minipage}{0.5\columnwidth}
\centerline{\includegraphics[width=1.3\columnwidth]{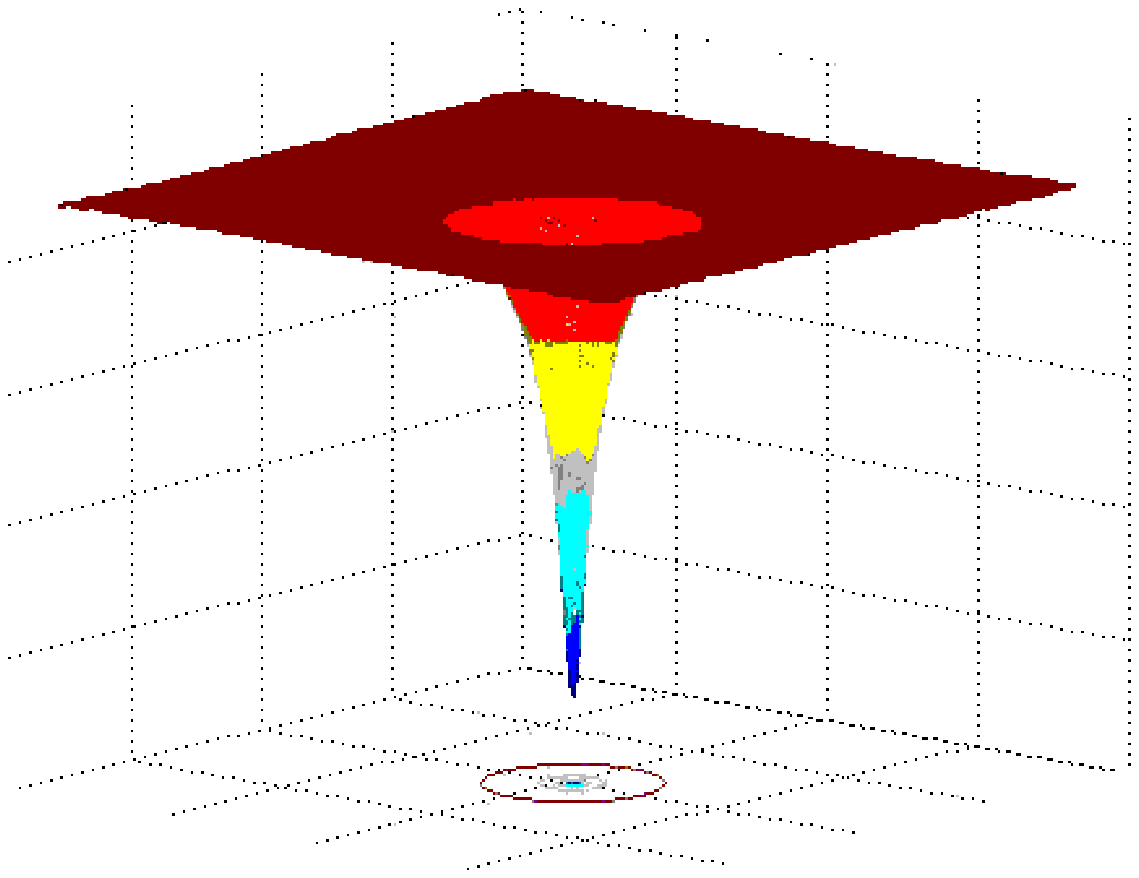}}
\end{minipage}
\caption{The $l$-plane case without bearing field, initial vortex,
$T=0$ and $T\approx 96 \rm h$. }\label{l_0_M0}
\end{figure}

\begin{figure}[h]
\begin{minipage}{0.5\columnwidth}
\centerline{\includegraphics[width=1.3\columnwidth]{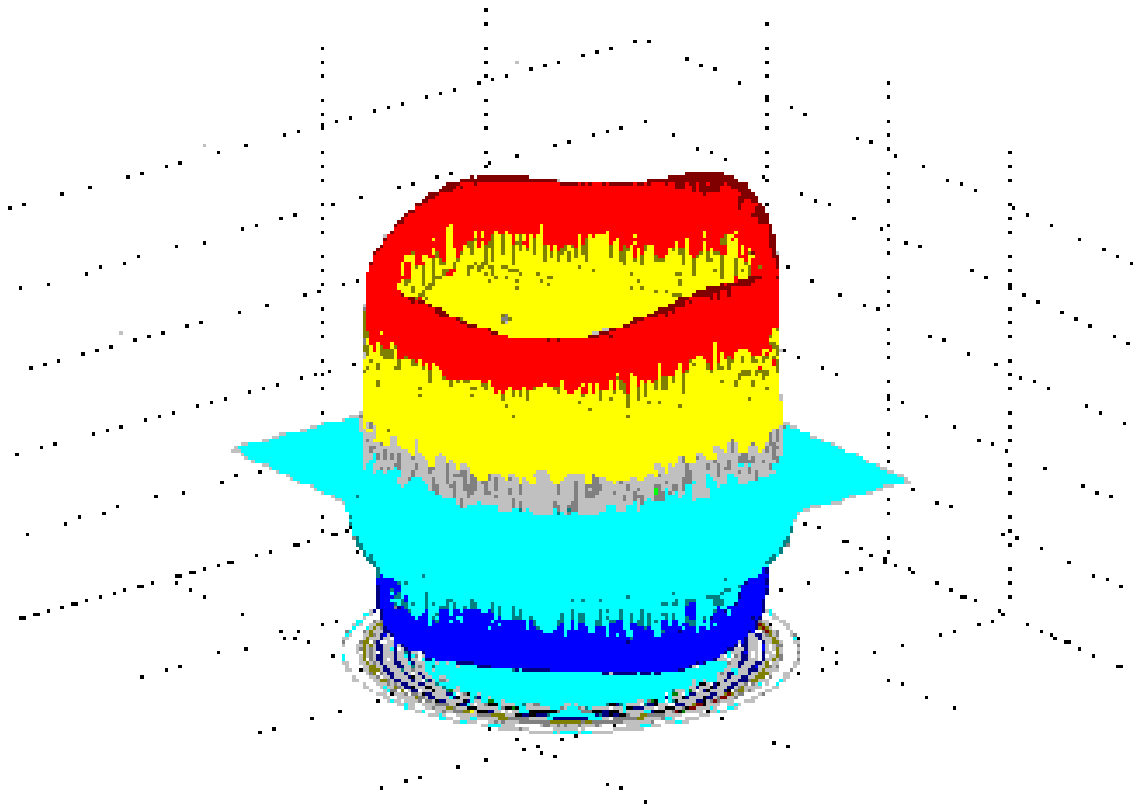}}
\end{minipage}%
\begin{minipage}{0.5\columnwidth}
\centerline{\includegraphics[width=1.3\columnwidth]{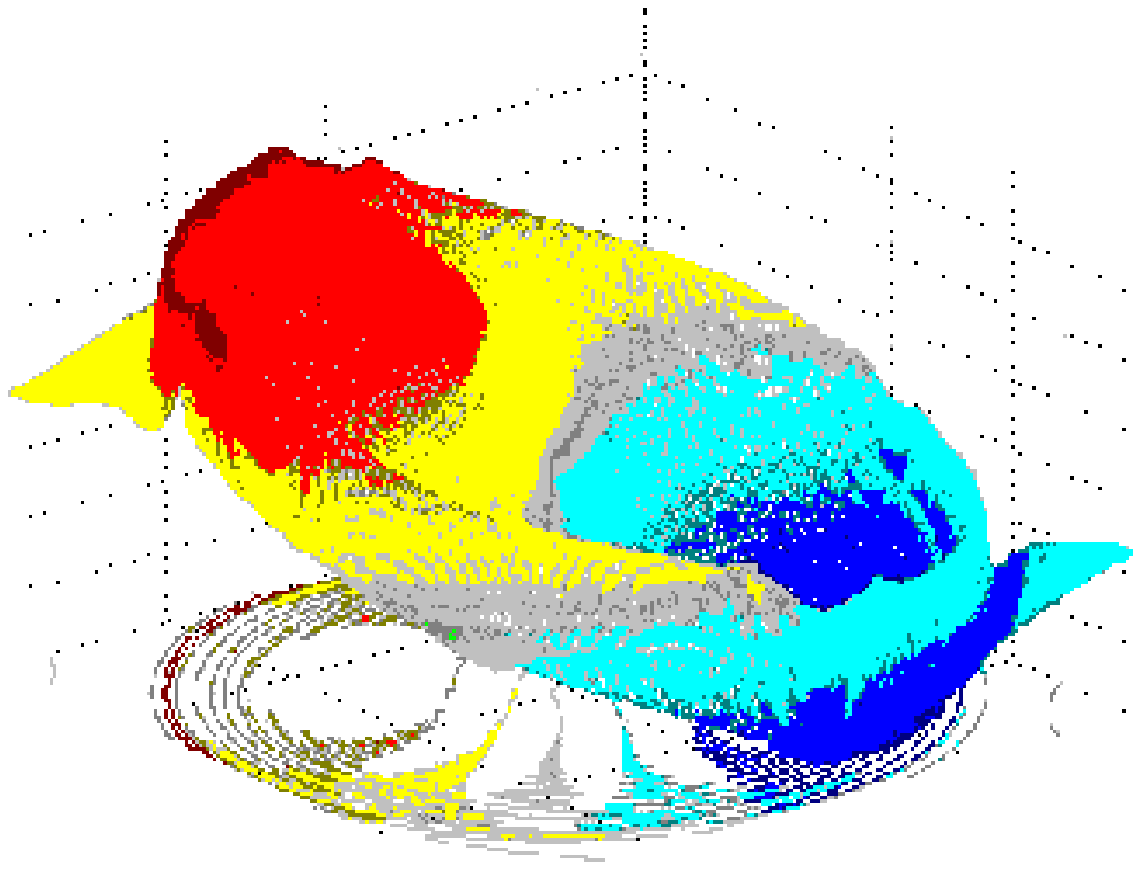}}
\end{minipage}
\caption{The pressure, weak and stronger bearing fields, $T\approx 1
\,\rm days$}\label{3Dpressure}
\end{figure}

\begin{figure}[h]
\begin{minipage}{0.5\columnwidth}
\centerline{\includegraphics[width=1.3\columnwidth]{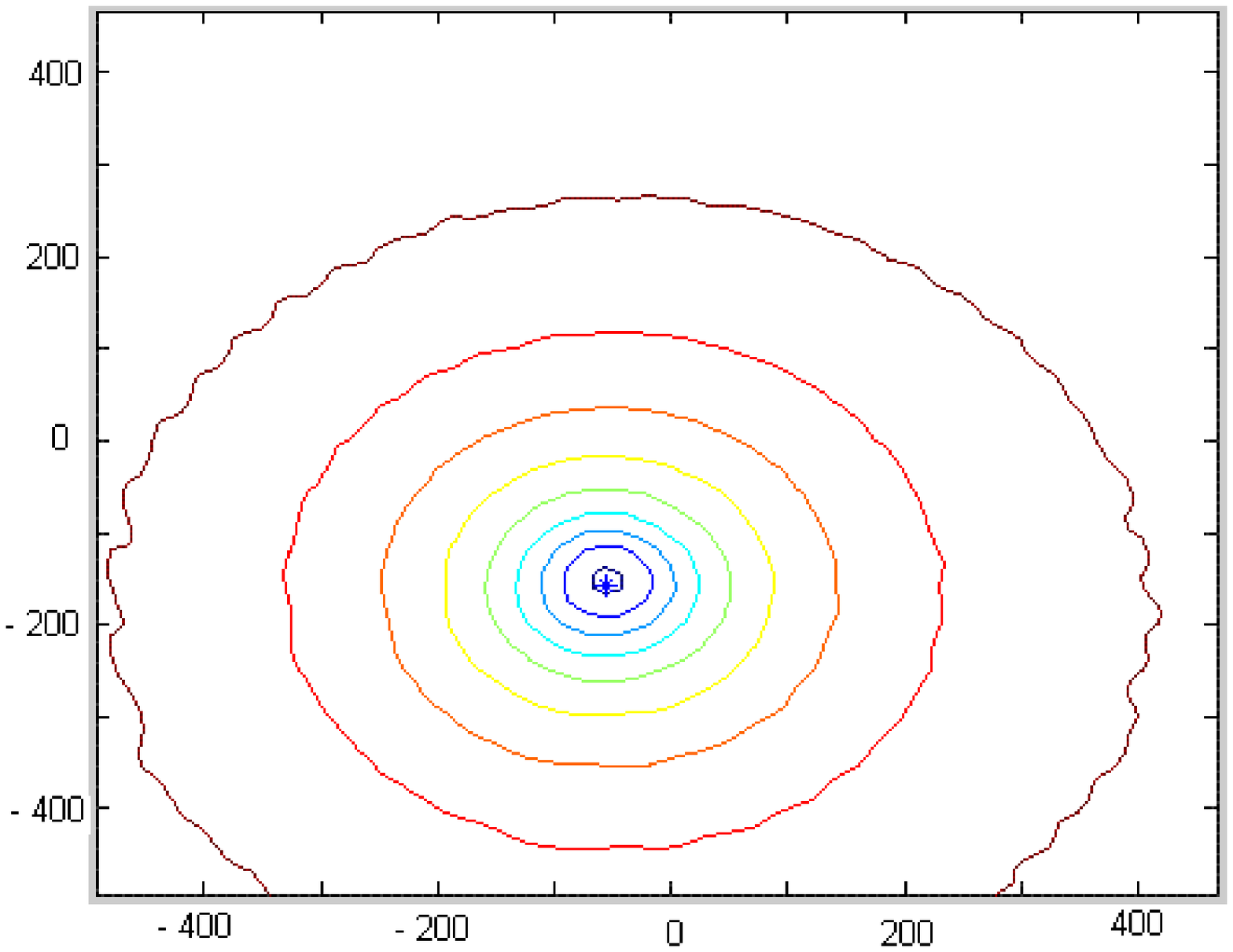}}
\end{minipage}
\begin{minipage}{0.5\columnwidth}
\centerline{\includegraphics[width=1.3\columnwidth]{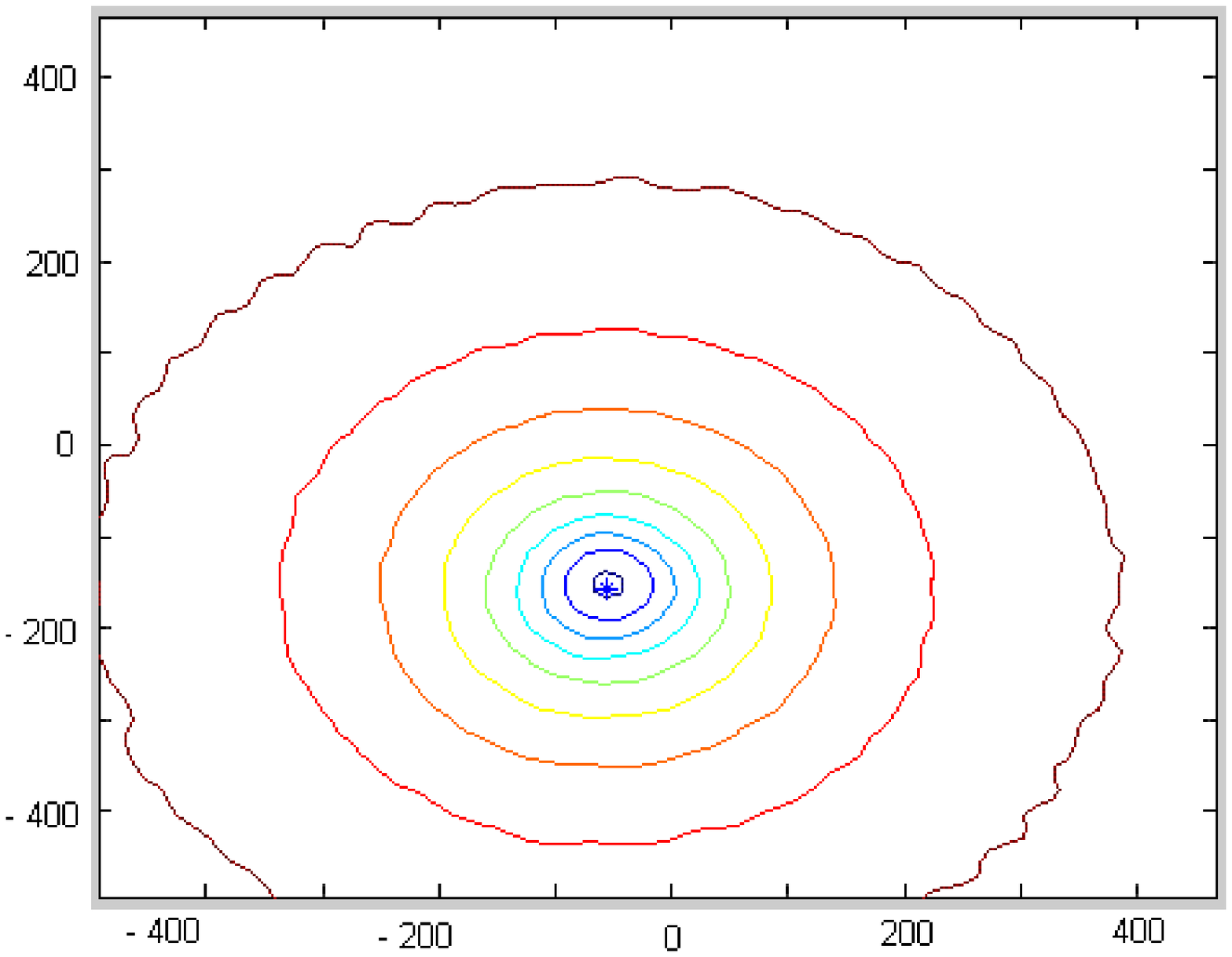}}
\end{minipage}%
\caption{The $l$-plane and the $\beta$ - plane cases, respectively,
for
 $T\approx 48 \rm h$, weak bearing
field.}\label{l_0_Mn0}
\end{figure}

Figs.\ref{l_0_M0} -- \ref{l_0_Mn0}
present the result of computations for the $l$ - plane  and the
$\beta$ - plane models. Here $\,\varphi_0\,=\,30^\circ,\,
l_0\,\approx\,7.3\times 10^{-5}\,{\rm s}^{-1}, \,\beta \,\approx\,
2\times 10^{-11} {\rm s}^{-1}, \, u_{10}=V_1(0)= u_{20}=V_2(0)=10
\,{\rm m/s} $. The initial position of  vortex is the origin. The
parameter $B_0,\,\sigma,\, c_0$ are as in the Sec.\ref{Comparison of
trajectories}. The localized vortex is given initially as
\eqref{pi_0_l_0},
\begin{equation}\label{u_l_0_12}
u_1\,=\,u_{10}\,+\,B_0\,\sigma\,x_2\,e^{-\frac{\sigma}{2}\, \left(
{x_1}^{2}+{x_2}^{2} \right)},\qquad u_2\,=\,\,u_{20}\,
-B_0\,\sigma\,x_1\,e^{-\frac{\sigma}{2}\, \left( {x_1}^{2}+{x_2}^{2}
\right)}.
\end{equation}
Fig.\ref{l_0_M0} corresponds to the  pressure for $\pi_1(0,x_1,
x_2)=0$ (zero bearing field and discrepancy), the first picture
presents the level lines for the initial data (common for all
figures), the second one corresponds to the vortex computed for
approximately 4 days.
Fig.\ref{3Dpressure} corresponds to the vortex on the $l$ - plane in
the bearing field. The parameters of "inclination" of the bearing
field are $M_{10}=-10^{-5},\, M_{20}=10^{-5}$  (weak bearing field)
for the first picture and $M_{10}=-10^{-5},\, M_{20}=10^{-5}$
(stronger bearing field) for the second picture. Computations are
made for one  day, the space scale is the same for both pictures in
Fig.\ref{3Dpressure} and for the second picture in Fig. \ref
{l_0_M0}. We can see that the vortex preserves its initial shape for
the null bearing field, it modifies its form and enlarges in a weak
bearing field and tends to disappear in a strong bearing field.
Fig. \ref{l_0_Mn0} corresponds to the initial bearing field of form
\eqref{pi_1_l_0} with $\sigma_0=10^{-13}$ (here and below the
dimension is appropriate) for the $l$ - plane and the $\beta$- plane
models. The parameters of "inclination" of the bearing field are
$M_{10}=-10^{-5},\, M_{20}=10^{-5}$  (weak bearing field,
approximately 2 days period).  The blue asterisk corresponds to the
theoretical position of vortex computed according to \eqref{x_l_loc}
and \eqref{beta_position} for $l$-plane and $\beta$-plane models,
respectively. The position of the center of vortex, obtained
numerically correlates precisely with the
 theoretical results.
The computations were made by a modified Lax-Wendroff scheme, the
method is second order accurate in both space and time  variables
\cite{Roache}, \cite{Zhang}, \cite{Lin}. We apply the scheme to the
system in the conservation form written compactly as follows:
$$\frac{\partial U}{\partial t} = \frac{\partial F}{\partial x}+\frac{\partial G}{\partial y}+S,$$
where
\[U =\left (
\begin{array}{c}
\varrho\\
\varrho u_1\\
\varrho u_2
\end{array}
\right)\]

\[F =\left (
\begin{array}{c}
-\varrho u_1\\
-\varrho u_1^2-\frac{9c0}{16}\varrho^{\frac{16}{7}}\\
-\varrho u_1 u_2
\end{array}
\right)\]

\[G =\left (
\begin{array}{c}
-\varrho u_2\\
-\varrho u_1 u_2\\
-\varrho u_2^2-\frac{9c0}{16}\varrho^{\frac{16}{7}}
\end{array}
\right)\]

\[S =\left (
\begin{array}{c}
0 \\
l u_2\\
-l u_1
\end{array}
\right)\]\\
Suppose that the solution domain in $2D$ is divided into rectangular
cells. Let $U^n_{i,j} = U((i-\frac{1}{2})\Delta
x,(j-\frac{1}{2})\Delta y,n\Delta t)$ be the value of $U$ in the
center of $(x_i,y_j)$ at the time level $t^n.$ Let us denote
\begin{eqnarray*}
\overline{f}_{i,j} &=&
\frac{1}{4}({f_{i-\frac{1}{2},j-\frac{1}{2}}+f_{i-\frac{1}{2},j+\frac{1}{2}}+f_{i+\frac{1}{2},j-\frac{1}{2}}+f_{i+\frac{1}{2},j+\frac{1}{2}}}),\\
(\delta_xf)_{i,j} &=& \frac{1}{\Delta
x}\overline{(f_{i+\frac{1}{2},j}-f_{i-\frac{1}{2},j})^y},\\
(\delta_yf)_{i,j} &=& \frac{1}{\Delta
y}\overline{(f_{i,j+\frac{1}{2}}-f_{i,j-\frac{1}{2}})^x},\\
\overline{f^x}_{i,j} &=&
\frac{1}{2}(f_{i+\frac{1}{2},j}+f_{i-\frac{1}{2},j}),\\
\overline{f^y}_{i,j} &=&
\frac{1}{2}(f_{i,j+\frac{1}{2}}+f_{i,j-\frac{1}{2}}).
\end{eqnarray*}
The explicit two-step scheme is given by
\begin{eqnarray*}
U^{n+\frac{1}{2}}_{i+\frac{1}{2},j+\frac{1}{2}}&=&\overline{U}^n_{i+\frac{1}{2},j+\frac{1}{2}}+\frac{\Delta
t}{2}(\delta_x F+\delta_y G+\overline{S})^n_{i+\frac{1}{2},j+\frac{1}{2}},\\
U^{n+1}_{i,j}&=& U^n_{i,j}+\Delta t(\delta_x F+\delta_y
G+\overline{S})^{n+\frac{1}{2}}_{i,j}.
\end{eqnarray*}
The computations  were performed on a $(240\times 240)$ uniform grid
with the space step $\Delta x = \Delta y = 0.64$ which corresponds
to 12.8 km and the time step $\Delta t = 0.0005$ which corresponds
to 10 sec of the real time.  We use the Neumann boundary condition
set sufficiently far from the vortex domain. Nevertheless, it is
possible to use more sophisticated non-reflecting boundary
conditions \cite{Givoli} and introduce an artificial viscosity  to
damp the oscillations \cite{Reutskiy}.

\section{Steady vortices on the sphere}\label{Sec5}

In Sections \ref{Sec2} and \ref{Sec3}
 we mentioned that  all equations,
e.g. \eqref{2d_U} -- \eqref{2d_p}, \eqref{2Dsys}, \eqref{2d_sh_u},
\eqref{2d_sh_pi}, can be  written in curvilinear coordinates, in
this case we mean by derivatives the covariant derivatives with
respect to the metric of a Riemannian manifold.

For example let us rewrite system \eqref{2d_U} -- \eqref{2d_rho}
avoiding the tensor notation in the spherical coordinates, making
clear the influence of curvature of the space, as this system
usually appears in the geophysical textbooks (e.g.\cite{Pedloski}):
\begin{equation*}
 \varrho\,\left(\frac{d U}{dt} - \left(2\Omega + \frac{U}{R \cos \tilde\phi}\right) \,V \sin \tilde\phi +
  \frac{1}{R \cos \tilde\phi}\,\frac{\partial
  P}{\partial\tilde\lambda}\right)=f_1,
\end{equation*}
\begin{equation*}
 \varrho\,\left(\frac{d V}{dt}+ \left(2\Omega + \frac{U}{R \cos \tilde\phi}\right)\, U \sin \tilde\phi +
  \frac{1}{R}\,\frac{\partial P}{\partial\tilde\phi}\right)=f_2,
\end{equation*}
\begin{equation*}
 \frac{\partial}{\partial t} (\varrho \, \cos \tilde\phi) +  \frac{\partial}{\partial
 \tilde\lambda} \left(\frac{\varrho U \cos\tilde\phi}{R}\right)+\frac{\partial}{\partial
 \tilde\phi} \left(\frac{\varrho V \cos\tilde\phi}{R}\right)=0,
\end{equation*}
where $\displaystyle \frac{d }{dt}= \frac{\partial}{\partial
t}+\frac{U}{R \cos \tilde\phi}\,\frac{\partial}{\partial
\tilde\lambda}+\frac{V}{R}\, \frac{\partial}{\partial \tilde\phi},$
$R$ is the radius of the Earth.

Here we use the established notation $(\tilde\lambda, \tilde\phi)$
for the geographical longitude and latitude, $\tilde\lambda\in
(-\pi, \pi),$ $\tilde\phi\in (-\frac{\pi}{2}, \frac{\pi}{2})$ (we
mark coordinates by the character  tilde to avoid confusion with the
moving coordinate system, where the latitude and longitude are
denoted as $(\lambda, \phi)$). Further, $(U,V)$ and $(f_1, f_2)$
stand for the components of velocity and  the vector of exterior
forces $F$, respectively. The Coriolis parameter is $l=2\,\Omega
\,\sin \tilde\phi$, $\Omega $ is again the vertical component of the
angular velocity of the Earth.

The master equation \eqref{master} now has the form (in the moving
coordinate system)
\begin{equation}\label{Sph_master}\begin{aligned}
\frac{\partial^2\Phi}{\partial \lambda \partial \phi}\left(
\big(\frac{\partial\Phi}{\partial
\phi}\big)^2-\frac{1}{\cos^2\phi}\big(\frac{
\partial\Phi}{\partial \lambda} \big)^2 \right)+\\
\left( \frac{1}{\cos^2\phi}\,\frac{\partial^2\Phi}{\partial
\lambda^2}-\frac{\partial^2\Phi}{\partial \phi^2}
\right)\frac{\partial\Phi}{\partial
\lambda}\,\frac{\partial\Phi}{\partial \phi}-\frac{\sin
\phi}{\cos^3\phi}\,\big(\frac{\partial\Phi}{\partial
\lambda}\big)^3=0,
\end{aligned}\end{equation}

\begin{figure}
\centerline{\includegraphics[width=0.4\columnwidth]{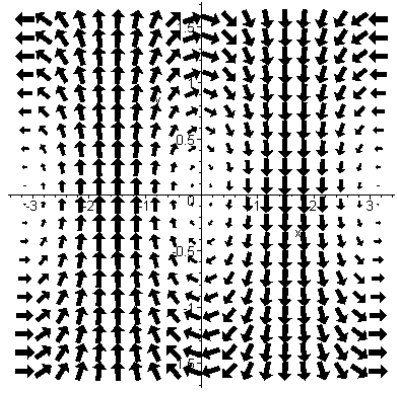}}
\caption{The periodic field of velocity corresponding to $\Phi=-\cos
\lambda \cos \phi$. The space scale is in radians.}\label{field_sh}
\end{figure}

Equation \eqref{Sph_master} has a solution of the form
\begin{equation}\label{Sph_phi}
\Phi(\lambda,\phi)=\Phi(\cos \phi (a\sin \lambda + b\cos
\lambda)+h\sin \phi),\,\mbox{where}\, a,b,h \,\mbox{are
constants}.\end{equation}
In particular, one can take
\begin{equation}
\Phi(\lambda,\phi)=b\cos \lambda \cos \phi,\quad b=\rm const,
\end{equation}
to obtain the zero of the respective velocity field at the origin of
the coordinate system. Thus, this field  is
\begin{equation}\label{vel_test}
u(\lambda,\phi)=-b\cos \lambda \sin \phi,\quad v(\lambda,\phi)=b\sin
\lambda.
\end{equation}
This field plays the similar  fundamental role in the case of the
spherical geometry  as the velocity with linear profile in the case
of the plane. It is easy to see that $(u,v)\sim (-b \phi, b\lambda)$
at the point $\lambda=\phi=0 $. To obtain a cyclonic (clockwise)
vorticity at the origin, we can set $b<0$. Fig.\ref{field_sh}
presents the field of velocity for $b=-1$. The velocity of form
\eqref{vel_test} was used for testing the numerical methods, e.g.
\cite{shallow_water},\cite{Jablonovski}.

If  $a^2+b^2\ne 0$, condition \eqref{cond_pi} can be satisfied only
for $A=l_0 L$, $l_0=\rm const$. For example, the  pressure field
$\pi_0$, corresponding to \ref{vel_test}, is
\begin{equation}
\pi_0(\lambda,\phi)=C-\frac{b\cos \lambda \cos\phi}{2
c_0}\left(b\cos\lambda \cos \phi + 2l_0\right),\, C=\rm const.
\end{equation}
The expansion at the origin has a form
\begin{equation}
\pi_0(\lambda,\phi)=C-\frac{b(b+l_0)}{2 c_0} - \frac{b(b+l_0)}{2
c_0} \left(\lambda^2+\phi^2\right) +
o(\left(\lambda^2+\phi^2\right)).
\end{equation}

Equation \eqref{pi1} can be also solved here; a rather cumbersome
solution is expressed in  elliptic functions $E_F(z,k)$ and
$E_{P}(z,\nu,k)$ (see \cite{Elliptic}):
\begin{equation}
\pi_1(t,\lambda,\phi)=G\left(\cos \phi \cos \lambda, b t +  {\cos
\phi \sin 2\lambda}\,\big(E_F(z,k)-2E_P(z,k,k)\big)\right),
\end{equation}
with arbitrary differentiable function $G$. Here
$$E_F(z,k)
=\int\limits_0^z\frac{1}{\sqrt{1-\xi^2}\sqrt{1-k^2\xi^2}}\,d\xi,$$
$$E_{P}(z,\nu,k)
      = \int\limits_0^z \frac{1}{(1-\nu\xi^2)\sqrt{1-\xi^2}\sqrt{1-k^2
      \xi^2}}\,
      d\xi,$$
      $$k=\nu=\frac{1+\cos \lambda \cos \phi}{1-\cos \lambda \cos \phi},\quad z=\frac{\cos \lambda -1}{\sin
      \lambda}\,\sqrt{k}.
      $$

It can be checked that for all $G(\eta_1,\eta_2)$ having bounded
partial derivatives for $\eta_1=0$ and all $\eta_2$
we have
$$\nabla\pi_1(t,\lambda, \phi)\Big|_{(\lambda, \phi)=0}=0.$$
 For the
sake of simplicity we take $\pi_1=0$. Then the trajectory of the
steady vortex can be obtained from \eqref{x}. In particular, for
$\Omega=0$ (non-rotating case)  and $l_0=0$ we get a stationary
exact solution $({\bf u}(\lambda,\phi), \pi_0(\lambda,\phi))$.

Since system \eqref{Sph_master}, \eqref{constraint} has many
solutions dependent on the choice of function $\Phi$, one can
construct a great variety of couples of "cyclonic" and
"anticyclonic" steady vortices on the non-rotating sphere.

If $\Omega\ne 0$ and/or $\pi_1\ne 0$, then the  position of the
respective   vortex can be found from  equation \eqref{x}:
\begin{equation}\begin{aligned} \label{xQ_sph}
\ddot{\bf X}(t)+ 2\Omega \sin ({\bf X}_2(t))\, L \,\dot{\bf
X}(t))+c_0 \nabla \pi_1(t,{\bf x})\Big|_{{\bf x}=0}= 0.\end{aligned}
\end{equation}
The discrepancy term for $A=2\Omega \sin X_2(0)$ is
\begin{equation}\begin{aligned}
Q-F =2\,\Omega \,\Big(\sin (\phi+X_2(t))\,-\,\sin X_2(t)\Big) \,L
\,\dot{\bf X}(t) +\\2\,\Omega\,(\sin(\phi+X_2(t))-\sin (X_2(0))
)\,L\,{\bf u}({\bf x})\,+\,c_0\,\left[\nabla \pi_1(t,{\bf
x})\Big|_{{\bf x}=0}- \nabla\pi_1(t,{\bf
x})\right],\end{aligned}\end{equation} here ${\bf
x}=(\lambda,\phi)$, ${\bf u}=(u,v)$. For $F=0$ it can be considered
as small for small $t$ in a small neighborhood of the center of the
theoretical vortex.

For $a=b=0$ in \eqref{Sph_phi} condition \eqref{cond_pi} is
satisfied for $A=2\Omega \sin \phi\, L$, the respective pressure
field  is
$$\pi_0(\lambda, \phi)\,=\,-\frac{1}{4}\,h\,(h+2\Omega)\,\sin 2\phi
\,+C.$$ Thus, we get an exact solution with $\pi_1=0$ for the
rotating sphere, moving according to \eqref{xQ_sph}. But evidently
it is not a vortex solution, it can be better interpreted as a zonal
flow.

We do not dwell here on the computations made for the case of  the
spherical geometry. This is a difficult issue, we reserve it for our
future research. Difference methods that can be useful here are
discussed in \cite{LeFloch}, \cite{Skiba}, \cite{Rossman}.

\section{Discussion}\label{Sec6}

 The vortex motion is intrinsic to the fluid, both compressible and
 incompressible. There is a huge literature, dedicated to the subject (e.g. \cite{Vortex1},\cite{Vortex2},\cite{Vortex3},
 \cite{VortexBibl}), especial emphases on vortex of rotating fluid and geophysical applications was made in \cite{Rotating}
 and \cite{Hopfinger}.
 In particular, the earliest results concerning vortices in the compressible fluid were obtained by C.Chree
 (e.g.\cite{chree}),
 it is remarkable that the author always bears in mind the meteorological context.
 Recently the vortices in compressible fluid were extensively
 studied by Shivamoggi (e.g.\cite{Shivamoggi}).

It is interesting that a very complicated systems of equations
describing the motion of compressible fluid  possesses  global in
time solutions of very simple form. For example, in the Euclidean
space it is a solution with linear profile of velocity, known since
Kirkhoff. Meanwhile the solution is not physically reasonable since
the velocity and the respective pressure field rise unboundedly as
the space variables go to infinity.

The present study was inspirited by a well known fact that the
tropical cyclones (typhoons) are  very stable atmospherical
structures and moves as a "rigid body" in a background field of
pressure. There were many attempts to use this fact for a
description of the trajectory of the vortex. It is possible to apply
a beautiful theory of point vortex (point singularity) at a sphere
\cite{bogomolov} to geophysical problems \cite{friedlander},
\cite{bogomolov1}, \cite{bogomolov2}. Further, there is a number of
paper, where the problem on the typhoon trajectory  was solved  in
assumption that the vortex is a square root singularity, e.g.
\cite{BVDD1}, \cite{DOR}. Nevertheless, the assumption on a singular
vortex structure contradicts to the data of observations \cite{IAV},
which evidence the linear profile of the velocity field near the eye
of typhoon.

 We propose a method of computing the
position of the center of vortex. The technique can be used for any
Riemannian manifold, nevertheless our main interest is a sphere and
a plane as a local flat approximation of the sphere near a point. We
consider two types of approximations: the $l$--plane model and the
$\beta$ -- plane model.

We solve a problem of finding the trajectory of a steady vortex
moving with a bearing field  of pressure. Namely, we consider a
special kind of solution to the system describing the motion of
compressible fluid on a rotating two-dimensional manifold.

Our method consists of several steps.

a) We use the fact that the vertical scale of the atmosphere is
small compared with the horizontal one. This allows to average the
whole atmosphere model over the height and to reduce it to two space
dimensions. The vertical structure plays the crucial role on the
phase of formation and maintenance of the stable vortex. The
averaging procedure hides these vertical processes, this simplifies
the model and helps to manage with the problem of the trajectory
describing. For the sake of simplicity we dwell on the  barotropic
case.

b) Further, as is known from experiments, near the eye of cyclone
the motion is axial-symmetric, the divergency vanishes and the
tangential component of velocity rises linearly (see \cite{IAV}). We
put the origin at the center of the vortex having coordinates
($X_1(t), X_2(t)$) (here we follow  \cite{BVDD1}) and in the new
coordinates ($t,x_1, x_2$) we perform a procedure that can be called
an approximate separation of variables. Namely, we separate all
terms into three parts: the first depends only on the new space
variables ($x_1, x_2$), the second depends only on $t$, and the
third depends both on space and time variables however from some
reasons can be considered as small. In this connection we use a
notion of the $\delta$-approximate solution of the reduced system of
atmospheric motion in a specific subdomain of the manifold where the
system is given.

c) Given a  vortex structure we find the stable component of
(re-normalized) pressure corresponding to the vortex. Further from a
linear PDE we find the bearing part of the pressure field. Then we
get a second order ODE for the vortex center position $X(t)$ (in
general case it includes the distance from the center as a small
parameter). In the case of a sphere we  find  exact stationary
solution of the planetary scale (non-rotating case) and study an
approximate vortex solution  in the rotative case.

We investigate the question on the difference between trajectory of
non-localized  vortex on the $l$ - plane where the exact solution
can be obtained and  the  trajectories of localized vortex on $l$-
plane and $\beta$ - plane that seems more realistic. We show that
the difference is reasonably small.

We do not stop here on the case of the presence of (turbulent)
viscosity in the model. This situation can be considered in our
framework. The difference is in the form of the master equation, in
other words, the problem is to find a stationary solution to the
Navier-Stokes equation. The class of such solution is not so rich as
in the non-viscous case, we refer for a comprehensive review of this
question to \cite{Wang}.



\begin{thebibliography}{99}




\bibitem{Elliptic}{\sc M.~Abramowitz,  I.~Stegun,} {\em Handbook of Mathematical
Functions.} Dover Publications Inc., New York, 1965.

\bibitem{Alishaev}{\sc  D.M.~Alishaev,} {\em On dynamics of two-dimensional baroclinic
atmosphere} \rm Izv.Acad.Nauk, Fiz.Atmos.Oceana,{\bf 16}(1980), N 2,
pp.~99-107.

\bibitem{LeFloch} {\sc M.Ben-Artzia, J.Falcovitza,  Ph.G.LeFloch,}
{\em Hyperbolic conservation laws on the sphere. A
geometry-compatible finite volume scheme,} Journal of Computational
Physics, 228 (2009), pp.5650-5668.

\bibitem{bogomolov} {\sc V.A.~Bogomolov},  {\em Dynamics of vorticity at a sphere} Translated from
Izvestiya Akademii Nauk SSSR, Mekhanika Zhidkosti i Gaza, No. 6,
1977, pp.~57--65.

\bibitem{bogomolov1}{\sc  V.A.~Bogomolov, }  {\em Two dimensional fluid dynamics on a
sphere}, Izv. Atmos. Ocean. Phys. 15(1979), pp.~18.

\bibitem{bogomolov2} {\sc V.A.~Bogomolov, }   {\em On the motion of a vortex on a rotating
sphere,} Izv. Atmos. Ocean. Phys. 21 (1985), pp.~298.

\bibitem{BVDD1}{\sc V.V.~Bulatov,  Yu.V.~Vladimirov,  V.G.~Danilov,
S.Yu.~Dobrokhotov,} {\em  On motion of the point algebraic
singularity for two-dimensional nonlinear equations of
hydrodynamics} Math. Notes 55 (1994), No.3, 243-250; translation
from Mat. Zametki 55, No.3, 11-20.

\bibitem{ChenWang}{\sc G.-Q.~Chen, D.~Wang} {\em The Cauchy problem for the Euler equation
s for compressible fluid} in: Handbook of Mathematical Fluid
Dynamics, Vol. 1, Elsevier, 2002.

\bibitem{chree}{\sc C.~Chree, } {\em Vortex Rings in a Compressible Fluid}, Proceedings of the Edinburgh Mathematical
Society, 6(1887), pp.~59-68, Cambridge University Press.

\bibitem{DOR}
{\sc V.~Danilov, G.~Omelyanov, D.~Rozenknop},    {\em Calculations
of the hurricane eye motion based on the singularity propagation
theory,} Electronic Journal of differential equations, No.16(2002),
pp.~1 -- 17.

\bibitem{DST}{\sc  S.Yu.~Dobrokhotov, A.I.~Shafarevich, B.~Tirozzi},
{\em The Cauchy-Riemann conditions and localized asymptotic
solutions of linearized equations in shallow water theory, }
Appl.Math.Mech. { 69}(2006) (5), pp.~720-725.

\bibitem{DKM}{\sc  F.V.~Dolzhanskii, V.A.~Krymov,  D.Yu.~Manin, }  {\em Stability and vortex structures of
quasi-two-dimensional shear flows} Sov.Phys.Usp., { 33}(1990), (7),
pp.~495-520.

\bibitem{Hopfinger}{\sc  E.J.~Hopfinger, G.J.F.~van Heijst, }  {\it Vortices in rotating
fluids}. Annual review of fluid mechanics,  25 (1993), pp.~241--289,
Annual Reviews, Palo Alto, CA.

\bibitem {Feireisl} {\sc  E.~Feireisl,} {\em Mathematical theory of viscous fluids: retrospective and future
perspectives.} Discrete Contin. Dyn. Syst.  27  (2010),  no. 2,
pp.~533--555.

\bibitem{friedlander}{\sc S.~Friedlander, }  {\em Interaction of vortices on the surface of a rotating sphere,}
Tellus 27(1975), pp~15-24.

\bibitem{Givoli} {\sc D.Givoli, B.Neta,}{\em High-order nonreflecting boundary conditions
for the dispersive shallow water equations,}
 J. Comput. Appl. Math. 158 (2003), pp.49-60.

\bibitem{Landau}{\sc  L.D.~Landau, E.M.~Lifshits, }
{\it Fluid mechanics.} 2nd ed. Volume 6 of Course of Theoretical
Physics., Oxford etc.: Pergamon Press. XIII, 1987.

\bibitem{shallow_water} {\sc D. Lanser, J. G. Blom, J. G. Verwer}, {\em Time Integration of the
Shallow Water Equations in Spherical Geometry}, Journal of
Computational Physics, 171(2001), pp.373-393.


\bibitem{Vortex1} {\sc C.C.~Lim, X.~Ding, J.~Nebus,  }{\em Vortex dynamics,
statistical mechanics, and planetary atmospheres.} World Scientific
Publishing Co. Pte. Ltd., Hackensack, NJ, 2009.

\bibitem{Lin}{\sc X.Lin, J.Ballmann,} {\em A numerical scheme for axisymmetric
elastic waves in solids }, Wave Motion,  21 (1995), pp.115-126.


\bibitem{Vortex3} {\sc  H.J.~Lugt,}{\em Introduction to vortex theory.} Vortex Flow Press,
Incorporated, Potomac, MD, 1996.


\bibitem{VortexBibl}{\sc V.V.~Meleshko,  H.~Aref,} {\em A Bibliography of Vortex Dynamics, \,1858 –- 1956}.  Advances
in Applied Mechanics, v.41 (2007), pp.~197-292

\bibitem{Jablonovski} {\sc R. Nair, C. Jablonowski,} {\em Moving vortices on the sphere: a test
case for horizontal advection problems,} Month. Weather Rev. 136
(2008), pp.699–711.

\bibitem{Obukhov}{\sc  A.M.~Obukhov,} {\em On the geostrophical wind}
\rm Izv.Acad.Nauk (Izvestiya of Academie of Science of URSS), Ser.
Geography and Geophysics,{ XIII}(1949), pp.~281-306.

\bibitem{Pedloski}{\sc J.Pedlosky, }  {\em Geophysical fluid dynamics},
Springer-Verlag, New York, 1979.

\bibitem{Reutskiy}{\sc S. Reutskiy,  B.Tirozzi} {\em Forecast of the trajectory of the center of typhoons and the Maslov
decomposition},  Russian Journal of Mathematical Physics, 14 (2007),
pp.232-237.

\bibitem{Roache} {\sc  P.J.Roache,} {\em Computational fluid dynamics},  Hermosa Publishers,
Albuquerque, N.M., 1976.

\bibitem{Rossman}{\sc J.A. Rossmanith,}{\em  A wave propagation method for
hyperbolic systems on the sphere,} Journal of Computational Physics,
V.213 (2006), pp.629-658.

\bibitem{OR2004} {\sc O.S.~Rozanova, }  {\em  Note on the typhoon eye
trajectory,} Regular and Chaotic Dynamics,  9 (2)(2004), pp.~129 --
142.






\bibitem{OR_JLY_CKH_2010}  {\sc O.S.~Rozanova,  J-L.~Yu,  C-K.~Hu,  }  {\em Typhoon
eye trajectory based on a mathematical model: Comparing with
observational data.}  Nonlinear Analysis: Real World Applications,
11 (2010),  pp.~1847--1861





\bibitem{Temam} {\sc M.~Petcu,  R.~Temam,  M.~Ziane,}  {\em Some mathematical problems in geophysical fluid
dynamics.} Handbook of numerical analysis. Vol.XIV.
Elsevier/North-Holland, Amsterdam, 2009.

\bibitem{Rotating}{\em  Rotating fluids in geophysical and industrial
applications.}, Edited by E.J.~ Hopfinger. CISM Courses and
Lectures, 329. Springer-Verlag, Vienna, 1992.

\bibitem{Vortex2} {\sc  P.G.~Saffman,}  {\em  Vortex dynamics}. Cambridge
Monographs on Mechanics and Applied Mathematics. Cambridge
University Press, New York, 1992.



\bibitem{IAV} {\sc R.C.~Sheets, } {\it  On the structure of hurricanes as revealed by research
aircraft data,} In: Intense atmospheric vortices. Proceedings of the
Joint Simposium (IUTAM/IUGC) held at Reading (United Kingdom) July
14-17, 1981. Edited by L.Begtsson and J.Lighthill, 33-49, 1981.



\bibitem{Shivamoggi} {\sc B.K.~Shivamoggi, }  {\em Vortex stretching and reconnection in a compressible
fluid,} The European physical journal. B, Condensed Matter Physics,
 49 (4)(2006), 483-490.

 \bibitem{Skiba} {\sc Yu.N.Skiba, D.M.Filatov,} {\em Conservative arbitrary order finite
difference schemes for shallow-water flows,} Journal of
Computational and Applied Mathematics Volume 218 (2008), pp.579-591.

\bibitem{Diff_Geom} {\sc M.~Spivak},   {\it  A comprehensive introduction to differential
geometry}. Vol. II. Second edition. Publish or Perish, Inc.,
Wilmington, Del., 1979. 

\bibitem{Wang}{\sc C.Y.~Wang }  {\it Exact solution of the steady-state Navier-Stokes equations}, Annu. Rev. Fluid Meeh.
23(1991), ~159-77.

\bibitem{Zhang} {\sc Y.Zhang,  B.Tabarrok,} {\em Modifications to the
Lax-Wendroff scheme for hyperbolic systems with source terms,}
Internat.J.Numer.Methods Engrg. 44 (1999), 27–40.

\end{thebibliography}
\end{document}